\documentclass{pasj00}
\draft
\Received{$\langle$reception date$\rangle$}
\Accepted{$\langle$acception date$\rangle$}
\Published{$\langle$publication date$\rangle$}
\SetRunningHead{Broad-line and Multi-wave Band Emission from
Blazars}{T. F. Yi, G. Z. Xie}

\usepackage{times}

\begin{document}

\title{Broad-line and Multi-wave Band Emission from Blazars}
\author{T. F. Yi,\altaffilmark{1}
        G. Z. Xie,\altaffilmark{2,3\dag}
        }
\altaffiltext{1}{Physics Department, Yunnan University, Kunming
650091, P. R. China} \altaffiltext{2}{National Astronomical
Observatories/Yunnan Observatory, Chinese Academy\\ of Sciences,
P. O. Box 110, Kunming 650011, P. R. China}
\altaffiltext{3}{Yunnan Astrophysics Center, Yunnan University,
Kunming 650091, P. R. China} \altaffiltext{\dag}{send offprint
requests to G. Z. Xie; Email:gzxie@public.km.yn.cn}

\maketitle

\begin{abstract}
We study the correlations of the flux of the broad-line emission
($F_{BLR}$) with the X-ray emission flux, optical emission flux at
5500 \AA\ and radio emission flux at 5 GHz, respectively, for a
large sample of 50 Blazars (39 flat-spectrum radio quasars (FSRQs)
and 11 BL Lac objects). Our main results are as follows. There are
very strong correlations between $F_{BLR}$ and $F_{X}$ and between
$L_{BLR}$ and $L_{X}$ in both states for 39 FSRQs and the slopes
of the linear regression equations are almost equal to 1. There
are weak correlations between $F_{BLR}$ and $F_{X}$ and between
$L_{BLR}$ and $L_{X}$ for 11 BL Lac objects in both states, and
the slopes of the linear regression equations are close to 1.
There are significant correlations between $F_{BLR}$ and $F_{X}$
and between $L_{BLR}$ and $L_{X}$ for 50 blazars in both states,
the slopes of both the linear regression equations are also close
to 1. These results support a close link between relativistic jets
and accretion on to the central Kerr black hole. On the other
hand, we find that BL Lac objects have low accretion efficiency
$\eta$, whereas FSRQs have high accretion efficiency $\eta$. The
unified model of FSRQs and BL Lac objects is also discussed.
\end{abstract}

\section{Introduction}

The relation between jets and accretion processes in active
galactic nuclei (AGN) is one of the most fundamental and open
problems. In current theoretical models of the formation of the
jet, the power is through accretion and then extracted from the
disk/black hole spin energy and converted in the kinetic power of
the jet (Blandford \& Znajek 1977; Blandford \& Payne 1982;
Maraschi \& Tavecchio 2003). In both cases the magnetic field play
a major role in channelling power from the black hole spin or disk
into the jet, in both scenarios it should be sustained by matter
accreting onto the black hole, leading one to expect a relation
between accretion power and jet power (Maraschi et al. 2003; Xie
et al. 2007). Recently, the concept of jet-disk symbiosis was
introduced (Cao \& Jiang 1999). The relativistic jet model plus
mass and energy conservation in the jet-disk system was applied to
study the relation between disk and jet luminosities (Cao \& Jiang
1999; Falcke \& Biermanm 1995; Falcke, Malkan \& Biemann 1995). An
effective approach to explore the link between the mentioned above
two phenomena is to study the relation between broad-line emission
luminosity and jet power at different wavelength (Dai et al. 2007;
Cao \& Jiang 1999; Maraschi \& Tavecchio 2003; Celotti, Padovani
\& Ghisellini 1997; Serjeant et al. 1998; Xu et al. 1999; Xie et
al. 2007). Celotti, Padovani \& Ghisellini (1997) considered a
large sample of radio-loud objects, and derived the accretion
luminosity from the Broad emission lines when available. They
explored the relation between the disk and jet using the
correlation between the broad-line luminosity and kinetic power.
The kinetic power of jets can be estimated by using the direct
radio emission data of the jets close the nucleus, as resolved by
very-long-baseline interferometry (VLBI) and the standard
synchrotron self-Compton theory of Celotti \& Fabian (1993). They
found evidence for a link between  jets and disks, although the
statistical  significance was too low to draw a firm conclusion.

Cao \& Jiang (1999) found a significant correlation between radio
and broad-line emission for a sample of radio-loud quasars that
supports a close link between accretion processes and relativistic
jets. It is well known that blazars are in fact the best
laboratories to study the physics of relativistic jets. For a
small sample of blazars, Maraschi \& Tavecchio (2003) also found a
significant correlation  between the luminosity carried by
relativistic jets and the nuclear luminosity provided by
accretion, using the good-quality broadband X-ray data provided by
the BeppoSAX satellite. Their results also support a close
accretion on to a rapidly spinning central black hole. However, in
a flux-limited sample that covers a wide range of redshift, a
correlation can appear in luminosity even though there is no
intrinsic correlation in the sources because the luminosity is
strongly correlated with redshift (M\"{u}cke et al. 1997). In
order to avoid the redshift bias to data, in this paper we will
discuss correlations between flux densities (and the corresponding
luminosities) in different wave bands because they are less
susceptible to such distortions.

In this paper, for a large sample of blazars, we study the
correlations of the broad-line flux with X-ray flux, optical flux
and radio flux, respectively, and the correlations of the
broad-line luminosity with X-ray luminosity, optical luminosity
and radio luminosity, respectively. The cosmological parameters
$H_{0}=75$ km s$^{-1}$ Mpc$^{-1}$ and $q_{0}=0.5$ have been
adopted in this work.

\section{Sample Description}
Based on the catalogue of blazars compiled by Donato et al. (2001)
and the catalogue of the broad emission line information of
radio-loud sources compiled by Cao \& Jiang (1999), we compiled a
sample of 50 blazars, for which the total broad-line flux (erg
cm$^{-2}$s$^{-1}$), X-ray flux at 1 keV in $\mu$Jy, optical flux
at 5500 \AA\ and radio flux at 5 GHz given by Cao \& Jiang (1999)
and Donato et al. (2001) are available. The relevant data for 50
blazars are listed in Table 1. The columns in this table are as
follows: Column (1): name of the source (IAU); column (2):
redshift; column (3): radio flux ($F_{R}$) at 5 GHz in Jy; column
(4): V band optical flux ($F_{O}$) in millijanskys (mJy); column
(5): 1 keV X-ray flux ($F_{X}$) in microjanskys ($\mu$Jy), the
first entry is for the high state, the second for the low state;
column (6): X-ray photon spectral index; column (7): estimated
total broad-line flux (erg cm$^{-2}$s$^{-1}$); Column (8): the
class. In Table 1, the data of X-ray, optical and radio bands are
taken from the catalogue compiled by Donato et al. (2001), and the
data of the broad-line emission are taken from the catalogue
compiled by Cao \& Jiang (1999). In table 1, for the X-ray band,
all 50 blazars have fluxes, but only 26 of them have fluxes in
both high and low states. For the remaining 24 objects, we can not
be certain whether they are in a high state or low state. In the
following analysis, we will consider two possible cases: (1)
assuming them to be in a high state, and (2) assuming them to be a
low state.

\section{Correlation Analysis of Fluxes and Luminosities between Various Wave Bands}
Radio, optical and X-ray flux densities are K-corrected according
to $F_{\nu}=F_{\nu}^{ob}(1+z)^{\alpha-1}$ and $\alpha_{R}=0.0$,
$\alpha_{O}=1.0$, $\alpha_{X}=\alpha_{X}^{ph}-1$ (Comastri et al.
1997). The X-ray photo spectral index $\alpha_{X}^{ph}$ has been
given by Donato et al. (2001) and listed in column (6) of Table 1.
Linear regression is applied to the relevant data to analyze the
correlations of flux densities and luminosities between different
wave bands. The analysis results are given in Table 2. The
principal results of
Table 2 are as follows:\\
(1) There is strong correlation between $\log F_{BLR}$ and $\log
\nu F_{X}$ in high state for 50 blazars (Fig. 1 and Table 2). The
linear regression equation in high state is
\begin{equation}
\label{eq:estmate_member} \log F_{BLR} =0.98\log \nu F_{X}-0.83,
\end{equation}
with correlation coefficient $r=0.64$ and chance probability
$p<10^{-4}$. The slope of the linear regression equation is almost
equal to 1. In Fig. 1, the solid line is the regression
line of all 50 blazars.\\
(2) From Table 2, one can see that there is very strong
correlations between $\log F_{BLR}$ and $\log \nu F_{X}$ in high
state for 39 FSRQs. The linear regression equation in high state
is
\begin{equation}
\label{eq:estmate_member}  \log F_{BLR}=1.00\log \nu F_{X}-0.43,
\end{equation}
with $r=0.76$ and $p<10^{-4}$. The result is also plotted in Fig.
1. The dash line is the linear regression equation (2). In
addition,
one notes that the slop of the equation (2) is almost equal to 1.\\
(3) There are weak correlations between $\log F_{BLR}$ and $\log
\nu F_{X}$ in both states for 11 BL Lac objects only (Table 2).
However, one can see that the slope of the linear regression
equation in high state is close to 1 (Table 2). The linear
regression equation in high state is
\begin{equation}
\label{eq:estmate_member}  \log F_{BLR}=0.74\log \nu F_{X}-4.59,
\end{equation}
with $r=0.65$ and $p=3.1\times10^{-2}$. The result is also plotted
in Fig. 1.
The dot line is the regression line of the BL Lac objects.\\
(4) A strong correlation ($r=0.67$, $p< 10^{-4}$) between $\log
F_{BLR}$ and $\log \nu F_{X}$ in low state has been found when
both FSRQs and BL Lac objects are considered (50 blazars). The
correlation is , however, much strong considering FSRQs only
($r=0.63$, $ p<10^{-4}$) and is weak for BL Lac objects only
($r=0.72$, $p=1.2\times10^{-2}$). The results are shown in Fig. 2.
In Fig. 2, the solid line is the regression line of all 50
blazars, the dash line and dot line are the regression lines of
FSRQs and BL Lac objects, respectively. From Table 2, one can see
that the slopes of the linear regression equations in low state
are close
to 1 in all cases.\\
(5) A significant correlation ($r=0.63$, $p<10^{-4}$) between
$\log F_{BLR}$ and $\log \nu F_{O}$ has been found when both FSRQs
and BL Lac objects are considered (50 blazars) (see Fig. 3). The
correlation is , however, very strong considering FSRQs only
($r=0.71$, $p<10^{-4}$) and is weak for BL Lac objects only
($r=0.68$, $p=2.2\times10^{-2}$). However, from Table 2 one can
see that the slopes of the linear
regression equations in optical band are close to 1 in all cases.\\
(6) A significant correlation ($r=0.43$, $p=1.9\times10^{-3}$)
between $\log F_{BLR}$ and $\log \nu F_{R}$ has been found when
both FSRQs and BL Lacs are considered (see Fig. 4). The
correlation is, however, very strong considering FSRQs only
($r=0.54$, $p\approx10^{-4}$)
and not for BL Lac objects only ($r=0.49$, $p=1.3\times10^{-1}$).\\
(7) A correlation between $\log L_{X}$ and $\log L_{BLR}$ for 50
blazars in high state (see Fig. 5) is also significant. The
correlation coefficient is $r=0.78$ and the chance probability is
$p< 10^{-4}$. The linear regression equation is
\begin{equation}
\label{eq:estmate_member}  \log L_{BLR}=1.23\log L_{X}-11.12.
\end{equation}
The slope of the linear regression equation (4) is close to 1.\\
(8) There is a very strong correlation between $\log L_{X}$ and
$\log L_{BLR}$ for 39 FSRQs in high state. The correlation
coefficient is $r=0.84$ and a chance probability is $p< 10^{-4}$.
The linear regression equation is
\begin{equation}
\label{eq:estmate_member}  \log L_{BLR}=1.12\log L_{X}-5.62.
\end{equation}
The slope of the linear regression equation (5) is very close to 1.\\
(9) There is a weak correlation between $\log L_{X}$ and $\log
L_{BLR}$ for 11 BL Lac objects in high state. The correlation
coefficient $r=0.73$ and the chance probability is
$p=1.0\times10^{-2}$. The linear regression equation is
\begin{equation}
\label{eq:estmate_member}  \log L_{BLR}=0.77\log L_{X}+8.95.
\end{equation}
The slope of the linear regression equation (6) is also close to
1.\\
(10) A strong correlation ($r=0.81$, $p< 10^{-4}$) between
 $\log L_{X}$ and $\log L_{BLR}$ in low state has been found for 50 blazars.
The correlations are significant for FSRQs only ($r=0.74$,
$p<10^{-4}$) and for BL Lac objects only ($r=0.79$,
$p=3.8\times10^{-3}$). As shown in Fig. 6, the solid line is the
regression line of all 50 blazars, and the dash line and dot line
are the regression lines of FSRQs and BL Lac objects,
respectively. In Table 2, We can see that the slopes of the linear
regression equations in low state are close to 1 in all cases.\\
(11) There are also strong correlations between $\log L_{O}$ and
$\log L_{BLR}$, and between $\log L_{R}$ and $\log L_{BLR}$, as
shown in Fig. 7 and 8, respectively. The slopes are also close to
1 in all cases.

\section{The Jet-Disk Relationship}
According to Ghisellini (2006), if relativistic jets are powered (
at least initially) by a Poynting flux, one can derive a simple
expression for the maximum value of their power, under some
reasonable assumptions. The Blandford \& Znajek (1977) power can
be written as:
\begin{equation}
\label{eq:estmate_member} L_{BZ}\sim6\times10^{20}
(\frac{a}{m})^2(\frac{M_{BH}}{M_{\odot}})^2B^{2}  \textrm{ erg}
\textrm{ s$^{-1}$},
\end{equation}
where $L_{BZ}$ is the luminosity, $M_{BH}$ is the mass of black
hole, $M_{\odot}$ is the solar mass, $\frac{a}{m}$ is the specific
black hole angular momentum($\sim 1$ for maximally rotating black
holes), and B is the magnetic strength in units of Gauss. Assuming
that the value of magnetic energy density $U_{B}\equiv
\frac{B^{2}}{8\pi}$ close to black hole is a fraction
$\varepsilon_{B}$ of the available gravitational energy:
\begin{equation}
\label{eq:estmate_member}
U_{B}=\varepsilon_{B}\frac{GM_{BH}\rho}{R}=\varepsilon_{B}\frac{R_{S}}{2R}\rho
c^{2},
\end{equation}
where $R_{S}=\frac{2GM_{BH}}{c^{2}}$ is the Schwarzschild radius.
Based on the accretion disk theory the mass density $\rho$ is
linked to the accretion rate $\dot{M}$  through
\begin{equation}
\label{eq:estmate_member} \dot{M}=2\pi RH \rho \beta_{r}c,
\end{equation}
where $\beta_{r}c$ is the radial in falling velocity, $H$ is the
disk thickness. As in stars, the fundamental process at work in an
active nucleus is the conversion of mass to energy. This is done
with some efficiency $ \eta $, the mass accretion rate
$\dot{M}=\frac{dM}{dt}$ is linked to the observed luminosity
produced by the accretion disk
\begin{equation}
\label{eq:estmate_member} L_{disk}=\eta\dot{M}c^{2}.
\end{equation}
From equations (7), (8), (9), (10), the Blandford-Znajek jet power
can be written as :
\begin{equation}
\label{eq:estmate_member}
L_{BZ,jet}\sim(\frac{a}{m})^{2}\frac{R_{S}^{3}}{HR^{2}}\frac{\varepsilon_{B}}{\eta}\frac{L_{disk}}{\beta_{r}}.
\end{equation}
The maximum jet power can be written as (Ghisellini 2006):
\begin{equation}
\label{eq:estmate_member} L_{jet}\sim\frac{L_{disk}}{\eta}.
\end{equation}
In addition, on the basis of present theories of accretion disk,
the broad-line region (BLR) is photoionized by a nuclear source
(probably radiation from the disk), Maraschi \& Tavecchio (2003)
obtained
\begin{equation}
\label{eq:estmate_member} L_{BLR}=\tau L_{disk},
\end{equation}
where $\tau$ represents the fraction of the central emission
reprocessed by the BLR, usually assumed to be 0.1. From equation
(12) and (13), we have
\begin{equation}
\label{eq:estmate_member} L_{BLR}=\tau\eta L_{jet}.
\end{equation}
It is well known that according to the accretion disk theory,
$\eta$ is not a constant. It ranges from 0.057 to 0.42. However in
our previous works (Xie et al. 2002; 2003), one prove that the
central black hole of blazars are probably Kerr black holes. Its
range should be narrow (Xie et al. 2005). From equation (14), we
can have
\begin{equation}
\label{eq:estmate_member} \log L_{BLR}=\log L_{jet}+\log
\eta+const.
\end{equation}
From the equation (15), we can see that the theoretical
predictions are as follows. (1) The slope of $\log L_{BLR}$-$\log
L_{jet}$ linear relation should be 1. $\eta$ is an important
physical parameter. (2) Since $\eta$ is not the same for all
objects, $\log \eta$ will contribute to the dispersion around the
linear relation.

\section{Discussions and Conclusions}
From Fig. 1 to Fig. 8 and Table 2, we can see that the straight
lines of linear regression equations of BL Lac parallel almost the
straight line of linear regression equations of FSRQs in X-ray,
optical and radio bands, respectively. However BL Lac objects
systematically lie somewhat below the FSRQs. On the other hand it
is worth noting that there is very strong correlation between
$F_{BLR}$ and $F_{X}$ in both states for all 50 Blazars, and there
are also significant correlations between $F_{BLR}$ and $F_{O}$ ,
and between $F_{BLR}$ and $F_{R}$. In addition, the slopes of the
straight lines of all linear regression equations of blazars in
X-ray, optical and radio bands are also close to 1. On the basis
of the theoretical results of equation (15) and the empirical
results mentioned above, we seem to obtain following implications.
(1) The coupling between the jet and accretion disk of blazars is
very close, leading to the fact that $F_{BLR}$ correlate
significantly with multi-wave band fluxes. (2) BL Lac objects are
a subclass of blazars which have low accretion coefficient $\eta$.
(3) FSRQs are also a subclass of blazars which have high accretion
coefficient $\eta$. (4) These results seem to show that FSRQs and
BL Lac objects should be regarded as a class which shows blazar
behavior (Xie et al. 2004; 2006). (5) The correlation between disk
and jet emission does not depend on the emission state of the AGN.
The idea of the unification of FSRQs and BL Lac objects is
confirmed by a new independent evidence. It is very interesting to
note that if we compare our results with the theoretical models of
Ghisellini (2006) and Maraschi \& Tavecchio (2003). We can see
that our experimental results are consistent with the theoretical
prediction of equation (15) in section 5 of this paper, this is,
these results mentioned above (both the experimental and the
theoretical) suggest a close link between the formation of
relativistic jet and accretion on to the central Kerr black hole.
Thus, we provide a solid basis for these theoretical speculations
of Ghisellini (2006) and Maraschi \& Tavecchio (2003).

From Table 2, one can see that there are very strong correlations
between  $F_{X}$ and $F_{BLR}$ or between $F_{O}$ and $F_{BLR}$ in
all cases. However, there is a significant correlation between
$F_{R}$ and $F_{BLR}$ for all blazars or for FSRQs, and  there is
weak correlation between $F_{R}$ and $F_{BLR}$ for BL Lac objects.
These experimental results support the theory that the optical-UV
and X-ray emission is energizing the emission of the broad-line
region, so the more optical-UV and X-rays emission in the core,
the more broad-line region emission. It is true that in blazars
the jet emission is highly beamed, so it is possible that much of
the emission of the observed X-ray flux may be from the jet
(although even in the brightest objects, the X-ray flux densities
are much weaker than the core). Even if the jet synchrotron
emission does dominate in the X-ray, to be called an FSRQ the
broad-line region emission must be commensurately bright enough to
be seen over the synchrotron continuum, thereby introducing these
correlations. This might explain why the correlations obtained in
this paper are weaker for BL Lac objects (because BL Lac objects
have no blue bump). In addition, the radio flux depends almost
exclusively on the parameters of a beamed jet, whereas the
optical-UV and X-ray flux depends on both the unbeamed accretion
disk and the beamed jet; and the broad-line region flux depend on
an unbeamed accretion disk. Therefore, these results lead us to
obtain that a strong correlation between $F_{X}$ and $F_{BLR}$ or
between $F_{O}$ and $F_{BLR}$, and a significant correlation
between $F_{R}$ and $F_{BLR}$, which based on the theoretical
model of the formation of the jet. It is these correlations and
difference of the correlations that imply a close link between
accretion disks and jets (i.e. jet-disk symbiosis).

The authors would like to thank the referee Elena Pian for her
helpful suggestions and comments.

\begin{figure}
\begin{center}
\includegraphics[angle=0,scale=1.5]{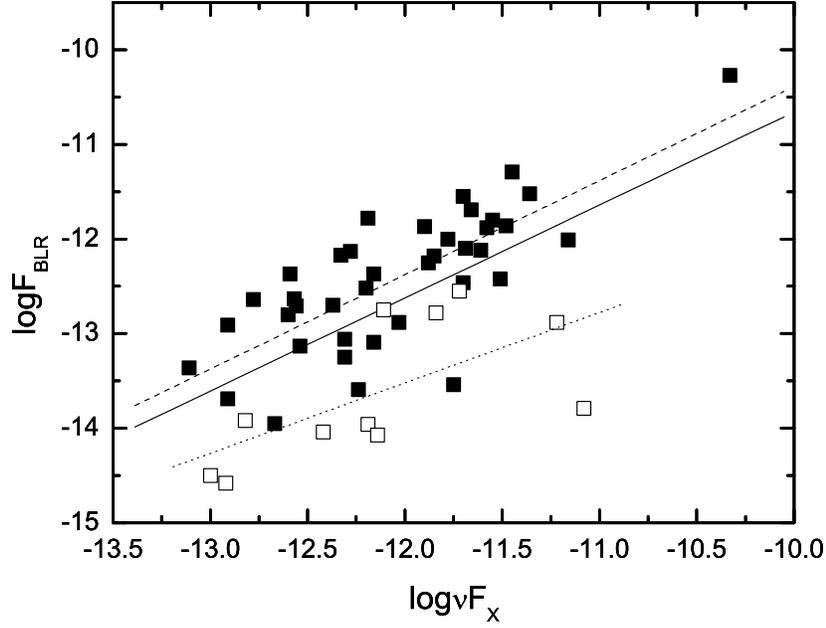}
\caption{Correlation between $\log \nu F_{X}$ and $\log F_{BLR}$
in high state for 50 blazars. The filled squares indicate 39
FSRQs, and the open squares indicate the 11 BL Lacs. The solid
line is the regression line of all 50 blazars ($k=0.98$, $r=0.64$
and $p<10^{-4}$), the dash line and dot line are the regression
lines of FSRQs ($k=1.00$, $r=0.76$ and $p<10^{-4}$) and BL Lacs
($k=0.74$, $r=0.65$ and $p=3.1\times10^{-2}$), respectively.}
\end{center}
\end{figure}
%\clearpage
\begin{figure}
\begin{center}
\includegraphics[angle=0,scale=1.15]{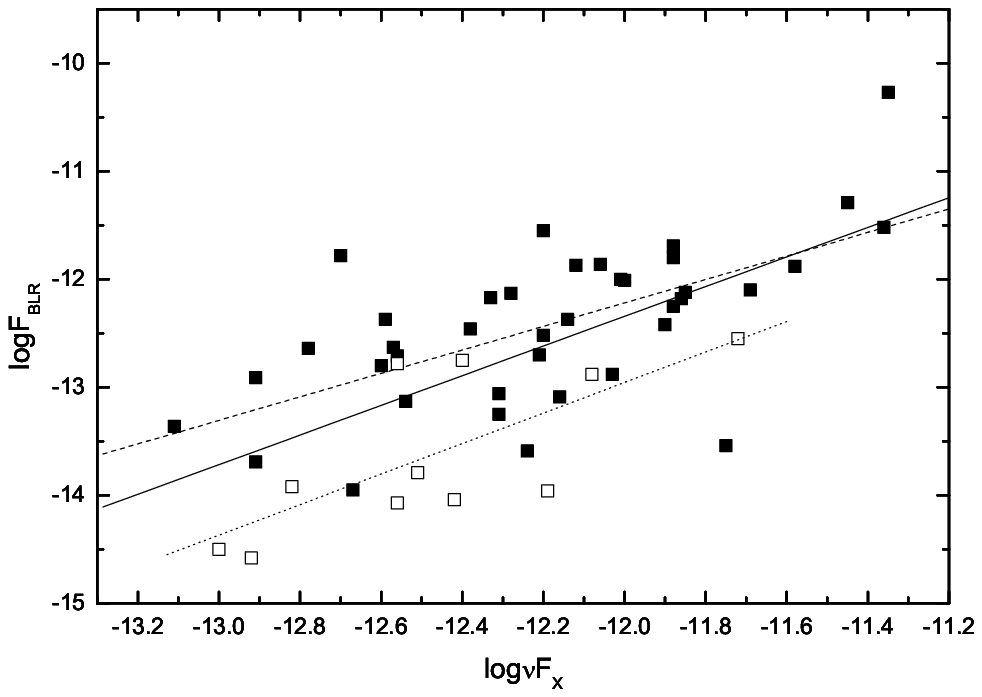}
\caption{Correlation between $\log \nu F_{X}$ and $\log F_{BLR}$
in low state for 50 blazars. The filled squares indicate 39 FSRQs,
and the open squares indicate the 11 BL Lacs. The solid line is
the regression line of all 50 blazars ($k=1.37$, $r=0.67$ and
$p<10^{-4}$), the dash line and dot line are the regression lines
of FSRQs ($k=1.09$, $r=0.63$ and $p<10^{-4}$) and BL Lacs
($k=1.41$, $r=0.72$ and $p=1.2\times10^{-2}$), respectively.}
\end{center}
\end{figure}
%\clearpage
\begin{figure}
\begin{center}
\includegraphics[angle=0,scale=1.2]{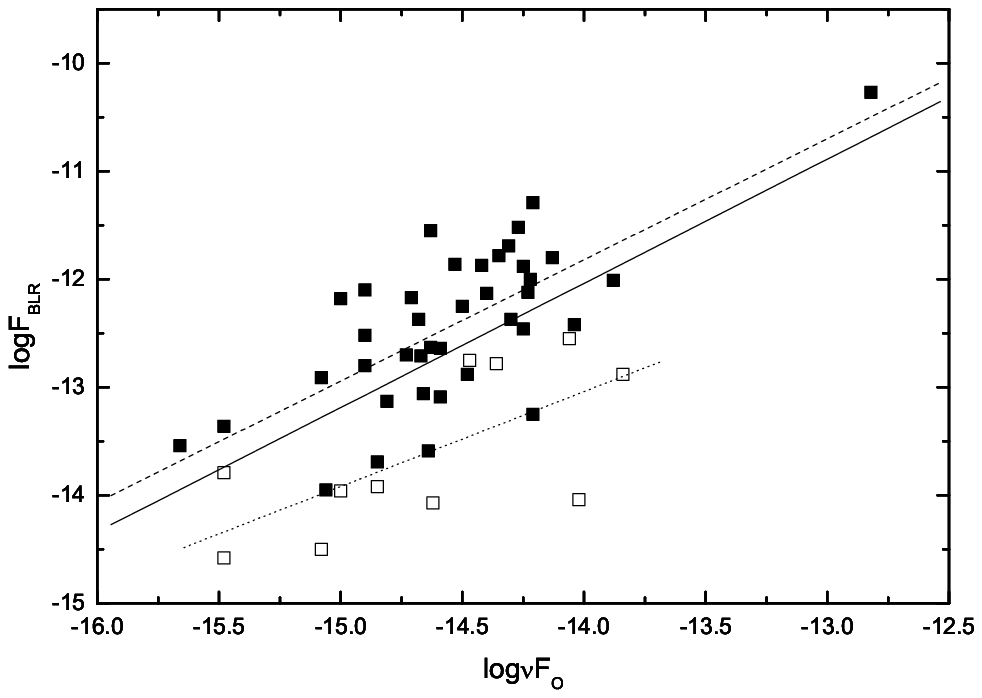}
\caption{Correlation between $\log \nu F_{O}$ and $\log F_{BLR}$
for 50 blazars. The filled squares indicate 39 FSRQs, and the open
squares indicate the 11 BL Lacs. The solid line is the regression
line of all 50 blazars ($k=1.15$, $r=0.63$ and $p<10^{-4}$), the
dash line and dot line are the regression lines of FSRQs
($k=1.12$, $r=0.71$ and $p<10^{-4}$) and BL Lacs ($k=0.88$,
$r=0.68$ and $p=2.2\times10^{-2}$), respectively.}
\end{center}
\end{figure}
%\clearpage
\begin{figure}
\begin{center}
\includegraphics[angle=0,scale=1.2]{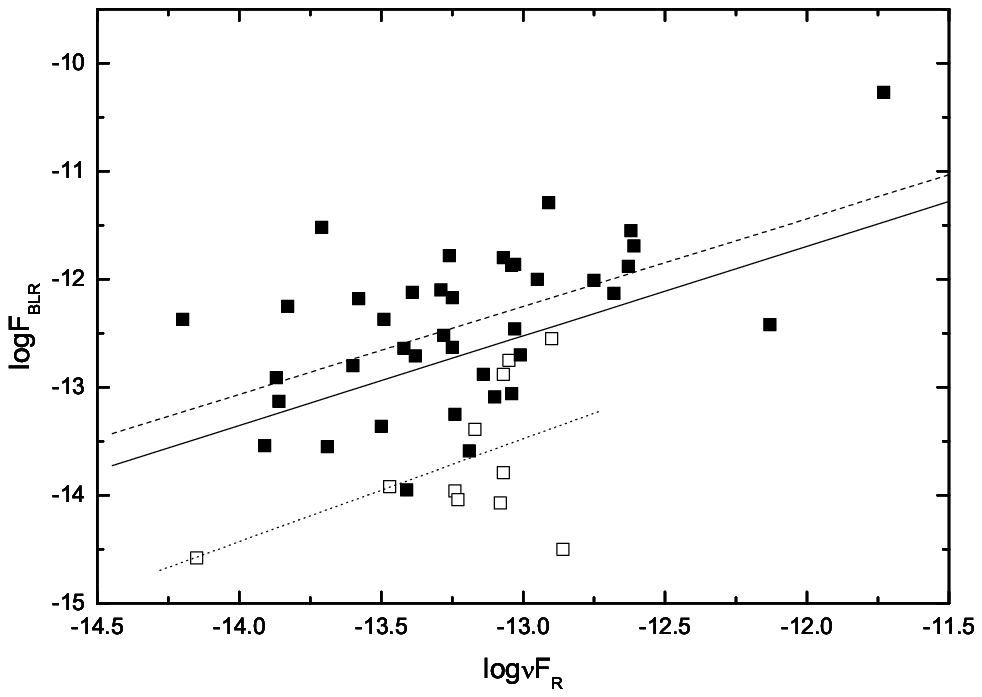}
\caption{Correlation between $\log \nu F_{R}$ and $\log F_{BLR}$
for 50 blazars. The filled squares indicate 39 FSRQs, and the open
squares indicate the 11 BL Lacs. The solid line is the regression
line of all 50 blazars ($k=0.83$, $r=0.43$ and
$p=1.9\times10^{-3}$), the dash line and dot line are the
regression lines of FSRQs ($k=0.81$, $r=0.54$ and
$p=3.9\times10^{-4}$) and BL Lacs ($k=0.95$, $r=0.49$ and
$p=1.3\times10^{-1}$), respectively.}
\end{center}
\end{figure}
%\clearpage
\begin{figure}
\begin{center}
\includegraphics[angle=0,scale=1.2]{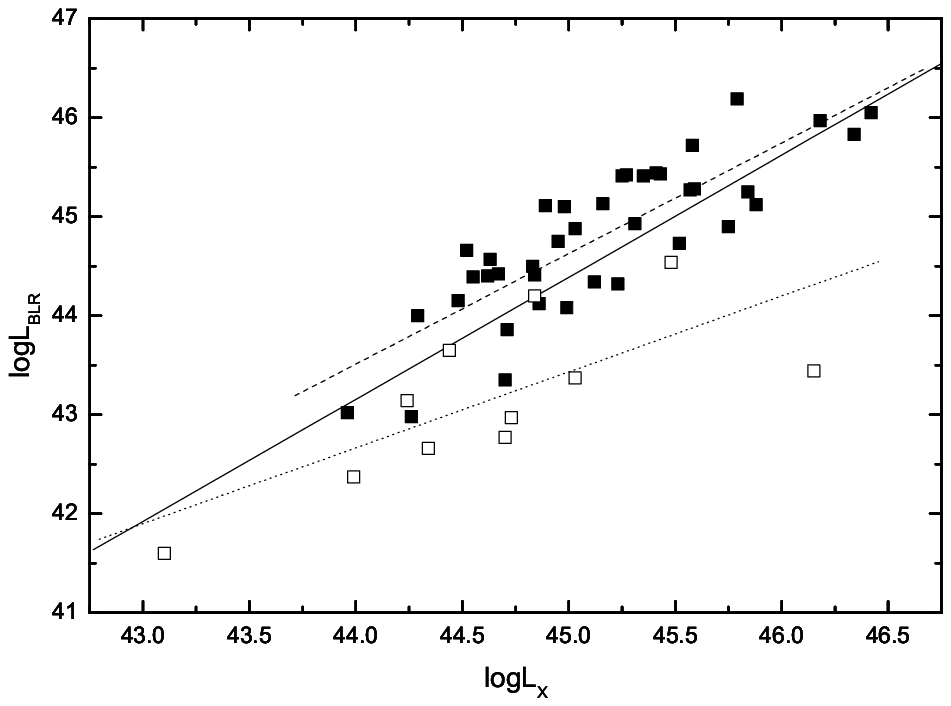}
\caption{Correlation between $\log L_{X}$ and $\log L_{BLR}$ for
50 blazars in high state. The filled squares indicate 39 FSRQs,
and the open squares indicate the 11 BL Lacs. The solid line is
the regression line of all 50 blazars ($k=1.23$, $r=0.78$ and
$p<10^{-4}$), the dash line and dot line are the regression lines
of FSRQs ($k=1.12$, $r=0.84$ and $p<10^{-4}$) and BL Lacs
($k=0.77$, $r=0.73$ and $p=1.0\times10^{-2}$), respectively.}
\end{center}
\end{figure}
%\clearpage
\begin{figure}
\begin{center}
\includegraphics[angle=0,scale=1.2]{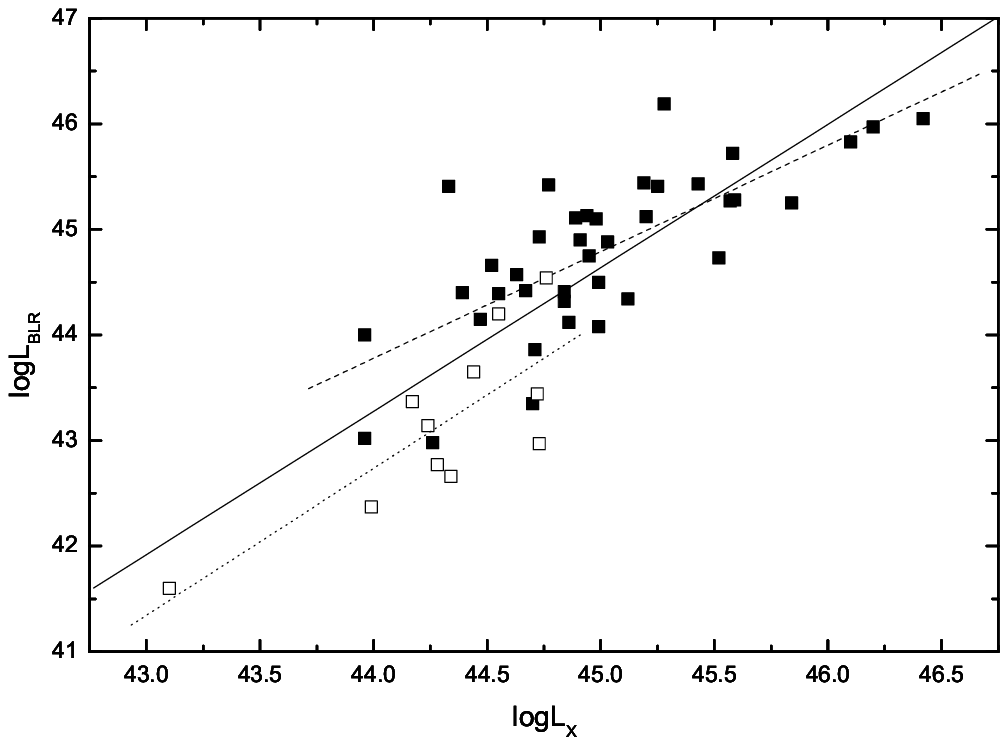}
\caption{Correlation between $\log L_{X}$ and $\log L_{BLR}$ for
50 blazars in low state. The filled squares indicate 39 FSRQs, and
the open squares indicate the 11 BL Lacs. The solid line is the
regression line of all 50 blazars ($k=1.36$, $r=0.81$ and
$p<10^{-4}$), the dash line and dot line are the regression lines
of FSRQs ($k=1.01$, $r=0.74$ and $p<10^{-4}$) and BL Lacs
($k=1.39$, $r=0.79$ and $p=3.8\times10^{-3}$), respectively.}
\end{center}
\end{figure}
\begin{figure}
\begin{center}
\includegraphics[angle=0,scale=1.2]{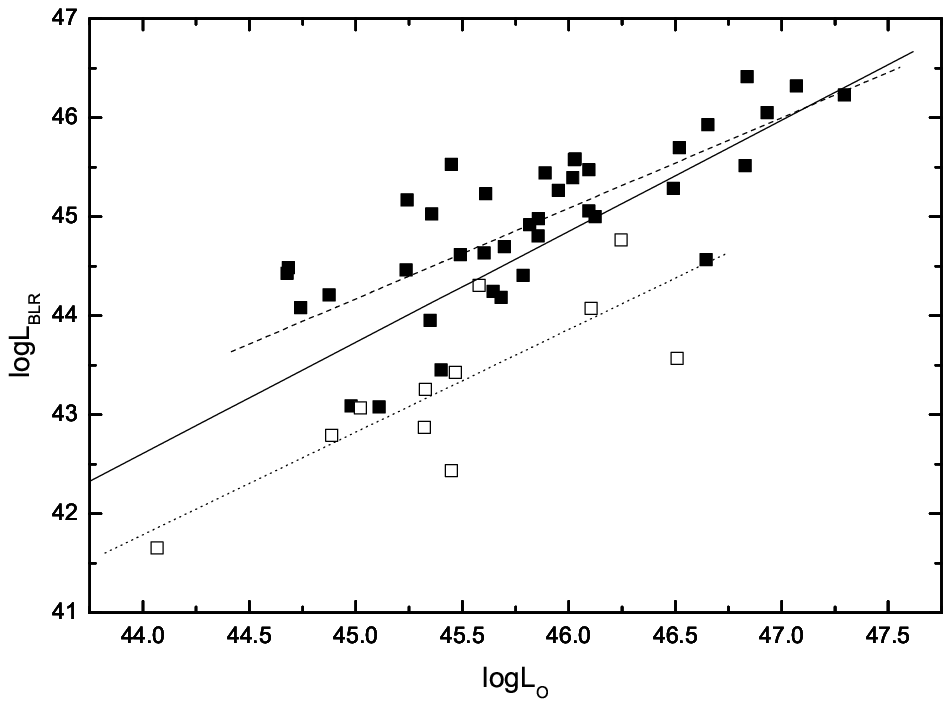}
\caption{Correlation between $\log L_{O}$ and $\log L_{BLR}$ for
50 blazars. The filled squares indicate 39 FSRQs, and the open
squares indicate the 11 BL Lacs. The solid line is the regression
line of all 50 blazars ($k=1.12$, $r=0.72$ and $p<10^{-4}$), the
dash line and dot line are the regression lines of FSRQs
($k=0.92$, $r=0.76$ and $p<10^{-4}$) and BL Lacs ($k=1.04$,
$r=0.80$ and $p=3.1\times10^{-3}$), respectively.}
\end{center}
\end{figure}
\begin{figure}
\begin{center}
\includegraphics[angle=0,scale=1.2]{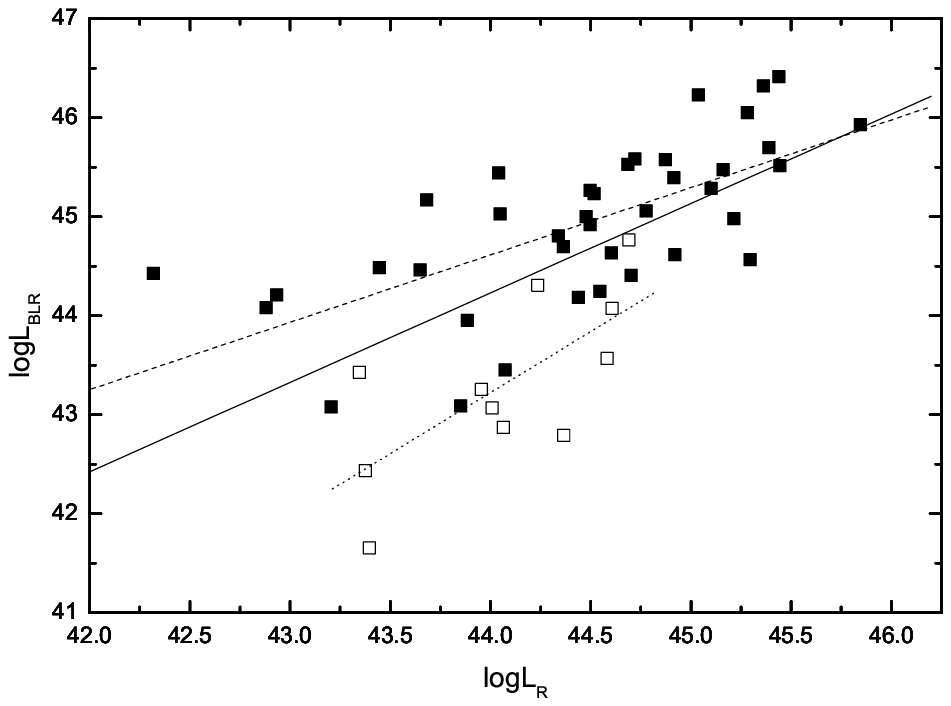}
\caption{Correlation between $\log L_{R}$ and $\log L_{BLR}$ for
50 blazars. The filled squares indicate 39 FSRQs, and the open
squares indicate the 11 BL Lacs. The solid line is the regression
line of all 50 blazars ($k=0.90$, $r=0.64$ and $p<10^{-4}$), the
dash line and dot line are the regression lines of FSRQs
($k=0.68$, $r=0.67$ and $p<10^{-4}$) and BL Lacs ($k=1.04$,
$r=0.70$ and $p=1.6\times10^{-2}$), respectively.}
\end{center}
\end{figure}

\begin{longtable}{cccccccl}
  \caption{Sample of 50 Blazars and the Relevant Data. Q: quasars; BL: BL Lac objects.}\label{tab:LTsample}
  \hline\hline
IAU Name &  z &  $F_{R}$  & $F_{O}$ &  $F_{X}$  &
 $\alpha$ & $\log F_{BLR}$   &  Class\\
& & (Jy)& (mJy)& ($\mu$Jy) &  & (erg cm$^{-2}$s$^{-1}$) &
\endfirsthead
\hline\hline
 IAU Name&z&  $F_{R}$  & $F_{O}$ &  $F_{X}$  &
 $\alpha$ & $\log F_{BLR}$   &  Class\\
 & & (Jy)&(mJy)& ($\mu$Jy) &  & (erg cm$^{-2}$s$^{-1}$) &\\
 \hline
\endhead
  \hline
\endfoot
  \hline
\endlastfoot
\hline
0014+813    &   3.366   &   0.551   &   0.91    &   0.43    &   0.72    &   -12.37  &   Q   \\
    &       &       &       &   0.35    &   0.89    &       &       \\
0016+731    &   1.781   &   1.750   &   0.06    &   0.05    &   0.43    &   -13.36  &   Q   \\
0112-017    &   1.365   &   1.200   &   0.23    &   0.15    &   0.57    &   -12.80  &   Q   \\
0133+476    &   0.859   &   2.920   &   0.47    &   0.30    &   0.92    &   -13.09  &   Q   \\
0212+735    &   2.370   &   2.198   &   0.45    &   0.26    &   -0.34   &   -13.69  &   Q   \\
0235+164    &   0.940   &   3.336   &   2.58    &   1.51    &   2.25    &   -13.79  &   BL  \\
    &       &       &       &   0.15    &   0.75    &       &       \\
0237-233    &   2.223   &   3.520   &   0.81    &   0.39    &   0.68    &   -11.78  &   Q   \\
    &       &       &       &   0.31    &   0.62    &       &       \\
0403-132    &   0.571   &   2.889   &   0.54    &   0.49    &   3.30    &   -11.86  &   Q   \\
    &       &       &       &   0.40    &   0.78    &       &       \\
0420-014    &   0.915   &   3.720   &   0.34    &   0.37    &   -0.15   &   -12.70  &   Q   \\
    &       &       &       &   0.28    &   0.85    &       &       \\
0537-286    &   3.104   &   0.990   &   0.04    &   0.18    &   0.36    &   -13.54  &   Q   \\
    &       &       &       &   0.17    &   0.50    &       &       \\
0537-441    &   0.896   &   4.755   &   1.57    &   0.78    &   1.04    &   -12.55  &   BL  \\
    &       &       &       &   0.17    &   0.27    &       &       \\
0637-752    &   0.651   &   5.849   &   2.38    &   3.76    &   0.45    &   -12.01  &   Q   \\
    &       &       &       &   0.49    &   0.64    &       &       \\
0642+449    &   3.400   &   1.204   &   0.15    &   0.12    &   0.41    &   -12.91  &   Q   \\
0736+017    &   0.191   &   1.999   &   1.33    &   1.49    &   1.82    &   -11.80  &   Q   \\
    &       &       &       &   0.37    &   3.20    &       &       \\
0804+499    &   1.433   &   2.050   &   0.39    &   0.17    &   0.56    &   -12.71  &   Q   \\
0814+425    &   0.258   &   3.500   &   0.15    &   0.05    &   0.16    &   -14.50  &   BL  \\
0820+225    &   0.951   &   1.977   &   0.06    &   0.05    &   1.05    &   -14.58  &   BL  \\
0836+710    &   2.170   &   2.590   &   1.06    &   2.26    &   0.31    &   -12.12  &   Q   \\
    &       &       &       &   1.27    &   0.32    &       &       \\
0851+202    &   0.306   &   2.173   &   2.64    &   2.24    &   1.37    &   -12.88  &   BL  \\
    &       &       &       &   0.37    &   0.71    &       &       \\
0906+430    &   0.670   &   1.300   &   0.16    &   0.11    &   0.57    &   -13.95  &   Q   \\
0923+392    &   0.699   &   8.101   &   0.43    &   0.88    &   0.88    &   -11.55  &   Q   \\
    &       &       &       &   0.37    &   0.36    &       &       \\
0945+408    &   1.252   &   1.450   &   0.38    &   0.11    &   0.96    &   -12.37  &   Q   \\
0954+556    &   0.909   &   2.169   &   0.43    &   0.10    &   1.17    &   -12.63  &   Q   \\
0954+658    &   0.367   &   1.589   &   1.72    &   0.16    &   0.96    &   -14.04  &   BL  \\
1034-293    &   0.312   &   1.510   &   1.11    &   0.24    &   0.41    &   -13.25  &   Q   \\
1040+123    &   1.029   &   1.560   &   0.47    &   0.1 &   0.45    &   -12.64  &   Q   \\
1055+018    &   0.888   &   3.470   &   0.40    &   0.21    &   0.93    &   -13.06  &   Q   \\
1144-379    &   1.048   &   2.779   &   0.80    &   0.41    &   1.54    &   -13.39  &   BL  \\
    &       &       &       &   0.14    &   0.70    &       &       \\
1150+497    &   0.334   &   0.699   &   0.560   &   0.63    &   0.77    &   -12.18  &   Q   \\
    &       &       &       &   0.55    &   1.14    &       &       \\
1226+023    &   0.158   &   42.861  &   27.64   &   20.42   &   0.51    &   -10.27  &   Q   \\
    &       &       &       &   1.96    &   0.57    &       &       \\
1253-055    &   0.536   &   14.950  &   1.66    &   1.50    &   0.65    &   -12.42  &   Q   \\
    &       &       &       &   0.63    &   0.55    &       &       \\
1334-127    &   0.539   &   2.250   &   0.60    &   0.45    &   0.63    &   -12.88  &   Q   \\
1442+101    &   3.530   &   1.260   &   0.28    &   0.12    &   1   &   -13.13  &   Q   \\
1510-089    &   0.361   &   3.080   &   1.09    &   0.83    &   0.38    &   -12.00  &   Q   \\
    &       &       &       &   0.49    &   0.35    &       &       \\
1538+149    &   0.605   &   2.648   &   0.44    &   0.28    &   1.13    &   -14.07  &   BL  \\
    &       &       &       &   0.09    &   1.05    &       &       \\
1546+027    &   0.413   &   1.450   &   0.23    &   0.84    &   1.18    &   -12.10  &   Q   \\
1611+343    &   1.404   &   2.671   &   0.35    &   0.24    &   0.76    &   -12.17  &   Q   \\
1633+382    &   1.814   &   2.929   &   0.23    &   0.42    &   0.53    &   -12.52  &   Q   \\
    &       &       &       &   0.25    &   0.51    &       &       \\
1641+399    &   0.594   &   7.821   &   0.89    &   0.98    &   0.85    &   -11.69  &   Q   \\
    &       &       &       &   0.70    &   0.43    &       &       \\
1721+343    &   0.206   &   0.468   &   0.98    &   1.99    &   0.50    &   -11.52  &   Q   \\
1803+784    &   0.684   &   3.016   &   0.61    &   0.26    &   1.42    &   -12.75  &   BL  \\
    &       &       &       &   0.22    &   0.45    &       &       \\
1823+568    &   0.664   &   1.905   &   0.18    &   0.41    &   0.15    &   -13.96  &   BL  \\
    &       &       &       &   0.27    &   0.96    &       &       \\
1928+738    &   0.360   &   3.339   &   1.11    &   1.47    &   1.33    &   -11.29  &   Q   \\
    &       &       &       &   0.52    &   1.25    &       &       \\
2126-158    &   3.268   &   1.240   &   0.58    &   0.86    &   0.68    &   -12.25  &   Q   \\
    &       &       &       &   0.71    &   2.18    &       &       \\
2134+004    &   1.936   &   12.249  &   0.72    &   0.26    &   0.82    &   -12.13  &   Q   \\
2155-152    &   0.672   &   2.150   &   0.42    &   0.22    &   1.17    &   -13.59  &   Q   \\
2223-052    &   1.404   &   4.519   &   1.02    &   1.46    &   0.34    &   -12.46  &   Q   \\
    &       &       &       &   0.22    &   0.73    &       &       \\
2230+114    &   1.037   &   3.689   &   0.69    &   0.73    &   0.51    &   -11.87  &   Q   \\
    &       &       &       &   0.42    &   0.58    &       &       \\
2240-260    &   0.774   &   1.203   &   0.26    &   0.07    &   0.79    &   -13.92  &   BL  \\
2251+158    &   0.859   &   8.759   &   1.02    &   1.37    &   0.62    &   -11.88  &   Q   \\
    &       &       &       &   1.31    &   0.83    &       &       \\

\end{longtable}

\begin{table*}
\caption{Linear Regression Analysis.} \small
 \begin{center}
   \begin{tabular}{cccccccl}\hline\hline
$X$ & $Y$ & $N$ & $A_{0}$& $k$&$r$ & $p$& class type \\
\hline
$\log \nu F_{X}$    &   $\log F_{BLR}$ &   50  &   -0.83  &   0.98   &   0.64   &   $< 10^{-4}$ &   Blazars \\
    &   (high state)  &       &       &       &       &       &       \\
$\log F_{X}$    &   $\log F_{BLR}$ &   39  &   -0.43   &   1.00   &   0.76   &   $< 10^{-4}$ &   FSRQ    \\
    &   (high state)  &       &       &       &       &       &       \\
$\log \nu F_{X}$    &   $\log F_{BLR}$ &   11  &   -4.59   &   0.74   &   0.65   &   3.1$\times10^{-2}$ &   BL Lac  \\
    &   (high state)  &       &       &       &       &       &       \\
$\log \nu F_{X}$    &   $\log F_{BLR}$ &   50  &   4.14   &   1.37   &   0.67   &   $< 10^{-4}$ &   Blazars \\
    &   (low state)   &       &       &       &       &       &       \\
$\log \nu F_{X}$    &   $\log F_{BLR}$ &   39  &   0.82   &   1.09   &   0.63   &   $< 10^{-4}$ &   FSRQ    \\
    &   (low state)    &       &       &       &       &       &       \\
$\log \nu F_{X}$    &   $\log F_{BLR}$ &   11  &   3.99    &   1.41   &   0.72    &   1.2$\times10^{-2}$ &   BL Lac \\
    &   (low state)    &       &       &       &       &       &       \\
$\log \nu F_{O}$   &   $\log F_{BLR}$ &   50  &   4.06   &   1.15    &   0.63   &   $< 10^{-4}$  &   Blazars \\
$\log \nu F_{O}$  &   $\log F_{BLR}$ &   39  &   3.89   &   1.12   &   0.71   &   $< 10^{-4}$ &   FSRQ    \\
$\log \nu F_{O}$   &   $\log F_{BLR}$ &   11  &   -0.74  &   0.88   &   0.68   &   2.2$\times10^{-2}$ &   BL Lac  \\
$\log \nu F_{R}$   &   $\log F_{BLR}$ &   50  &   -1.72  &   0.83   &   0.43   &   1.9$\times10^{-3}$ &   Blazars    \\
$\log \nu F_{R}$   &   $\log F_{BLR}$ &   39  &   -1.69  &   0.81   &   0.54   &   3.9$\times10^{-4}$ &   FSRQ    \\
$\log \nu F_{R}$   &   $\log F_{BLR}$ &   11  &   -1.11  &   0.95   &   0.49   &   1.3$\times10^{-1}$ &   BL Lac    \\
$\log L_{X}$    &   $\log L_{BLR}$ &   50  &   -11.12   &   1.23   &   0.78   &   $< 10^{-4}$   &   Blazars \\
    &  (high state) &       &       &       &       &       &       \\
$\log L_{X}$    &   $\log L_{BLR}$ &   39  &   -5.62   &   1.12   &   0.84   &   $< 10^{-4}$   &   FSRQ    \\
    &   (high state)  &       &       &       &       &       &       \\
$\log L_{X}$    &   $\log L_{BLR}$ &   11  &   8.95   &   0.77   &   0.73   &   1.0$\times10^{-2}$   &   BL Lac \\
    &   (high state)  &       &       &       &       &       &       \\
$\log L_{X}$    &   $\log L_{BLR}$ &   50  &   -16.51   &   1.36   &   0.81   &   $< 10^{-4}$   &   Blazars \\
    &  (low state) &       &       &       &       &       &       \\
$\log L_{X}$    &   $\log L_{BLR}$ &   39  &   -0.62   &   1.01   &   0.74   &   $< 10^{-4}$   &   FSRQ    \\
    &   (low state)  &       &       &       &       &       &       \\
$\log L_{X}$    &   $\log L_{BLR}$ &   11  &   -18.45   &   1.39   &   0.79   &   3.8$\times10^{-3}$   &   BL Lac \\
    &   (low state)  &       &       &       &       &       &       \\
$\log L_{O}$    &   $\log L_{BLR}$ &   50  &   -6.78   &   1.12   &   0.72   &   $< 10^{-4}$   &   Blazars \\
$\log L_{O}$    &   $\log L_{BLR}$ &   39  &    3.02   &   0.92   &   0.76   &   $< 10^{-4}$   &   FSRQ    \\
$\log L_{O}$    &   $\log L_{BLR}$ &   11  &   -3.83   &   1.04   &   0.80   &   3.1$\times10^{-3}$   &   BL Lac \\
$\log L_{R}$    &   $\log L_{BLR}$ &   50  &   4.49    &   0.9    &   0.64   &   $< 10^{-4}$   &   Blazars \\
$\log L_{R}$    &   $\log L_{BLR}$ &   39  &   14.65   &   0.68   &   0.67   &   $< 10^{-4}$   &   FSRQ    \\
$\log L_{R}$    &   $\log L_{BLR}$ &   11  &   -11.02  &   1.23   &   0.70   &   1.6$\times10^{-2}$   &   BL Lac \\

\hline
   \end{tabular}
\begin{flushleft}
Note: The linear regression is obtained by considering $X$ to be
the independent variable and assuming a relation $Y=kX+A_{0}$; $N$
is the number of points, $r$ is the correlation coefficient, and
$p$ is the chance probability.
\end{flushleft}
 \end{center}
\end{table*}

\end{document}